\documentclass[cup7b]{cupbook}

\begin{document}
%


\pagenumbering{roman}




\def\Box{{\hbox{$\sqcup$}\llap{\hbox{$\sqcap$}}}}

\def \lsim{\mathrel{\vcenter
     {\hbox{$<$}\nointerlineskip\hbox{$\sim$}}}}
\def \gsim{\mathrel{\vcenter
     {\hbox{$>$}\nointerlineskip\hbox{$\sim$}}}}

\cleardoublepage
\pagenumbering{arabic}

\author[C.P. Burgess]{C.P. BURGESS\\Department of Physics \& Astronomy,
        McMaster University,\\ 1280 Main St. W,
        Hamilton, Ontario, Canada, L8S 4M1\\ {\it and} \\
        Perimeter Institute for Theoretical Physics,\\
        31 Caroline St. N,
        Waterloo, Ontario, Canada, N2L 2Y5.}

\chapter{Quantum Gravity and Precision Tests} 

\begin{abstract}
This article provides a cartoon of the quantization of General
Relativity using the ideas of effective field theory. These ideas
underpin the use of General Relativity as a theory from which
precise predictions are possible, since they show why quantum
corrections to standard classical calculations are small. Quantum
corrections can be computed controllably provided they are made
for the weakly-curved geometries associated with precision tests
of General Relativity, such as within the solar system or for
binary pulsars. They also bring gravity back into the mainstream
of physics, by showing that its quantization (at low energies)
exactly parallels the quantization of other, better understood,
non-renormalizable field theories which arise elsewhere in
physics. Of course effective field theory techniques do {\it not}
solve the fundamental problems of quantum gravity discussed
elsewhere in these pages, but they do helpfully show that these
problems are specific to applications on very small distance
scales. They also show why we may safely reject any proposals to
modify gravity at {\it long} distances if these involve low-energy
problems (like ghosts or instabilities), since such problems are
unlikely to be removed by the details of the ultimate
understanding of gravity at microscopic scales.
\end{abstract}

\medskip
\begin{center}
{\it Contribution to `Towards Quantum Gravity,' edited by D.
Oriti, Cambridge University Press, 2006.}
\end{center}

\section{Introduction}

Any of us who has used the Global Positioning System (GPS) in one
of the gadgets of everyday life has also relied on the accuracy of
the predictions of Einstein's theory of gravity, General
Relativity (GR). GPS systems accurately provide your position
relative to satellites positioned thousands of kilometres from the
earth, and their ability to do so requires being able to
understand time and position measurements to better than 1 part in
$10^{10}$. Such an accuracy is comparable to the predicted
relativistic effects for such measurements in the Earth's
gravitational field, which are of order $G M_\oplus/R_\oplus c^2
\sim 10^{-10}$, where $G$ is Newton's constant, $M_\oplus$ and
$R_\oplus$ are the Earth's mass and mean radius, and $c$ is the
speed of light. GR also does well when compared with other precise
measurements within the solar system, as well as in some
extra-solar settings \cite{CWLivRev}.

So we live in an age when engineers must know about General
Relativity in order to understand why some their instruments work
so accurately. And yet we also are often told there is a crisis in
reconciling GR with quantum mechanics, with the size of quantum
effects being said to be infinite (or --- what is the same
--- to be unpredictable) for gravitating systems. But since precision
agreement with experiment implies agreement within both
theoretical and observational errors, and since uncomputable
quantum corrections fall into the broad category of (large)
theoretical error, how can uncontrolled quantum errors be
consistent with the fantastic success of classical GR as a
precision description of gravity?

This chapter aims to explain how this puzzle is resolved, by
showing why quantum effects in fact {\it are} calculable within
GR, at least for systems which are sufficiently weakly curved (in
a sense explained quantitatively below). Since all of the extant
measurements are performed within such weakly-curved environments,
quantum corrections to them can be computed and are predicted to
be fantastically small. In this sense we quantitatively understand
{\it why} the classical approximation to GR works so well within
the solar system, and so why in practical situations quantum
corrections to gravity need not be included as an uncontrolled
part of the budget of overall theoretical error.

More precisely, the belief that quantum effects are incalculable
within GR arises because GR is what is called a non-renormalizable
theory. Non-renormalizability means that the short-wavelength
divergences --- which are ubiquitous within quantum field theory
--- cannot be absorbed into the definitions of a finite number of
parameters (like masses and charges), as they are in
renormalizable theories like Quantum Electrodynamics (QED) or the
Standard Model (SM) of the strong and electroweak interactions.
Although this does preclude making quantum predictions {\it of
arbitrary accuracy}, it does not preclude making predictions to
any finite order in an appropriate low-energy expansion, and this
is what allows the predictivity on which precise comparison with
experiment relies. In fact gravity is not at all special in this
regard, as we know of other non-renormalizable theories which
describe nature --- such as the chiral perturbation theory which
describes the low-energy interactions of pions and kaons, or the
Fermi theory of the weak interactions, or a wide variety of
condensed matter models. In many of these other systems quantum
corrections are not only computable, they can be measured, with
results which agree remarkably well with observations.

One thing this chapter is {\it not} intended to do is to argue
that it is silly to think about the problems of quantum gravity,
or that there are no interesting fundamental issues remaining to
be addressed (such as many of those described elsewhere in these
pages). What is intended is instead to identify more precisely
where these more fundamental issues become important (at very
short distances), and why they do not hopelessly pollute the
detailed comparison of GR with observations. My presentation here
follows that of my longer review of ref.~\cite{CBLivRev}, in which
the arguments given here are provided in more detail.

\section{Nonrenormalizability and the low-energy approximation}

Since the perceived difficulties with calculating quantum
corrections in weak gravitational fields revolve around the
problem of calculating with non-renormalizable theories, the first
step is to describe the modern point of view as to how this should
be done. It is convenient to do so first with a simpler toy model,
before returning to GR in all of its complicated glory.

\subsection{A toy model}

Consider therefore the theory of a complex scalar field, $\phi$,
described by the Lagrangian density
\begin{equation}
\label{abeltoymodel}
 {\cal L} = - \partial_\mu \phi^*
 \partial^\mu \phi - V(\phi^* \phi) \,,
\end{equation}
with the following scalar potential
\begin{equation}
\label{toypotential}
 V = {\lambda^2 \over 4} \; \left( \phi^* \phi - v^2 \right)^2 \,.
\end{equation}
This theory is renormalizable, so we can compute its quantum
implications in some detail.

Since we return to it below, it is worth elaborating briefly on
the criterion for renormalizability. To this end we follow
standard practice and define the `engineering' dimension of a
coupling as $p$, where the coupling is written as (mass)${}^p$ in
units where $\hbar = c = 1$ (which are used
throughout).\footnote{It is implicit in this statement that the
relevant fields are canonically normalized, and so have
dimensionless kinetic terms.} For instance the coupling
$\lambda^2$ which pre-multiplies $(\phi^* \phi)^2$ above is
dimensionless in these units, and so has $p = 0$, while the
coupling $\lambda^2 v^2$ pre-multiplying $\phi^*\phi$ has $p = 2$.

A theory is renormalizable if $p \ge 0$ for all of its couplings,
and if for any given dimension all possible couplings have been
included consistent with the symmetries of the theory. Both of
these are clearly true for the Lagrangian of
eqs.~(\ref{abeltoymodel}) and (\ref{toypotential}), since all
possible terms are written consistent with $p \ge 0$ and the
$U(1)$ symmetry $\phi \to e^{i\omega} \phi$.

\subsubsection{Spectrum and scattering}

We next analyze the spectrum and interactions, within the
semiclassical approximation which applies in the limit $\lambda
\ll 1$. In this case the field takes a nonzero expectation value,
$\langle \phi \rangle = v$, in the vacuum. The particle spectrum
about this vacuum consists of two weakly-interacting particle
types, one of which -- $\varphi_0$ -- is massless and the other --
$\varphi_m$ -- has mass $m = \lambda v$. These particles interact
with one another through an interaction potential of the form
\begin{equation}
\label{potl} V = \frac12 \left[ m \varphi_m +
\frac{\lambda}{2\sqrt2} \Bigl( \varphi_m^2 + \varphi_0^2 \Bigr)
\right]^2,
\end{equation}
as may be seen by writing $\phi = v + (\varphi_m + i
\varphi_0)/\sqrt2$. For instance, these interactions imply the
following invariant scattering amplitude for the scattering
process $\varphi_0 (p) \varphi_0 (q) \to \varphi_0 (p') \varphi_0
(q')$
\begin{eqnarray}
\label{smatrixresult}
 {\cal A} &=& - \; {3 \lambda^2 \over 2} +
 \left( { \lambda m \over \sqrt2}
 \right)^2 \left[ {1 \over (p+q)^2 + m^2
 - i\epsilon}  \right. \nonumber\\
 && \qquad\qquad \left. + {1 \over (p-p')^2 + m^2 -i \epsilon}
 + {1 \over (p - q')^2 + m^2 - i\epsilon}\right] \,.
\end{eqnarray}
This amplitude has an interesting property in the limit that the
center-of-mass scattering energy, $E$, is much smaller than the
mass $m$. As may be explored by expanding ${\cal A}$ in powers of
external four-momenta, in this limit the $O(\lambda^2)$ and
$O(\lambda^2 E^2/m^2)$ terms both vanish, leaving a result ${\cal
A} = O(\lambda^2 E^4/m^4)$. Clearly the massless particles
interact more weakly than would be expected given a cursory
inspection of the scalar potential, eq.~(\ref{potl}).

The weakness of the scattering of $\varphi_0$ particles at low
energy is a consequence of their being Nambu-Goldstone bosons
\cite{ChiPT,CBGB} for the theory's $U(1)$ symmetry: $\phi \to
e^{i\omega} \phi$. This can be seen more explicitly by changing
variables to polar coordinates in field space, $\phi = \chi \;
e^{i \theta}$, rather than the variables $\varphi_0$ and
$\varphi_m$. In terms of $\theta$ and $\chi$ the action of the
$U(1)$ symmetry is simply $\theta \to \theta + \omega$, and the
model's Lagrangian becomes:
\begin{equation}
\label{linpolcoords} {\cal L} = - \partial_\mu \chi \partial^\mu
\chi - \chi^2 \partial_\mu \theta
\partial^\mu \theta - \frac{\lambda^2}{4} (\chi^2 - v^2)^2 \,,
\end{equation}
and semiclassical calculations can be performed as before by
expanding in terms of canonically-normalized fluctuations: $\chi =
v + \hat\varphi_m/\sqrt2$ and $\theta = \hat\varphi_0/v\sqrt2$,
revealing that $\hat\varphi_m$ describes the massive particle
while $\hat\varphi_0$ describes the massless one. Because
$\hat\varphi_0$ appears in ${\cal L}$ only explicitly
differentiated (as it must because of the symmetry $\hat\varphi_0
\to \hat\varphi_0 + \omega v \sqrt2$), its scattering is
suppressed by powers of $E/m$ at low energies.

\subsubsection{The low-energy effective theory}

Properties such as this which arise (sometimes unexpectedly) when
observables are expanded at low energies in powers of $E/m$ are
explored most easily by `integrating out' the heavy particle to
construct the {\it effective field theory} describing the
low-energy dynamics of the massless particle alone. One way to do
so in the case under consideration here would be to define `light'
degrees of freedom to be those modes (in momentum space) of
$\hat\varphi_0$ which satisfy $p^2 < \Lambda^2$ (in Euclidean
signature), for some cutoff $\Lambda$ satisfying $E \ll \Lambda
\ll m$. All other modes are, by definition, `heavy'. Denoting the
heavy and light modes schematically by $h$ and $\ell$, then the
effective theory governing the light fields may be defined by
\begin{equation}
\label{leffdef} \exp\left[i \int d^4x \; {\cal L}_{\rm
eff}(\ell,\Lambda) \right] = \int {\cal D}h_\Lambda \; \exp\left[
i \int d^4x \; {\cal L}(\ell,h) \right] \,,
\end{equation}
where the functional integral is performed over all of the heavy
modes (including the large-momentum components of
$\hat\varphi_0$).

${\cal L}_{\rm eff}$ defined this way necessarily depends on
$\Lambda$, but it does so in just the way required in order to
have $\Lambda$ cancel with the explicit $\Lambda$'s which cut off
the loop integrals for the functional integration over the light
fields, $\ell$. All $\Lambda$'s must cancel in observables because
$\Lambda$ is just a bookmark which we use to organize the
calculation. Because of this cancellation the detailed form of the
regularization is largely immaterial and can be chosen for
convenience of calculation.

For this reason it is actually preferable instead to define ${\cal
L}_{\rm eff}$ using dimensional regularization rather than a
cutoff. Paradoxically, this is possible even though one keeps both
short- and long-wavelength modes of the light fields in the
low-energy theory when dimensionally regularizing, which seems to
contradict the spirit of what a low-energy effective theory is. In
practice it is possible because the difference between the cutoff-
and dimensionally-regularized low-energy theory can itself be
parameterized by an appropriate choice for the effective couplings
within the low-energy theory. This is the choice we shall make
below when discussing quantum effects within the effective theory.

With this definition, physical observables at low energies are now
computed by performing the remaining path integral over the light
degrees of freedom only, weighted by the low-energy effective
Lagrangian: $\exp\left[i \int d^4x \; {\cal L}_{\rm eff}(\ell)
\right]$. The effects of virtual contributions of heavy states
appear within this low-energy theory through the contributions of
new effective interactions. When applied to the toy model to
leading order in $\lambda$ this leads to the following result for
${\cal L}_{\rm eff}$:
\begin{eqnarray}
\label{Leffform}
 {\cal L}_{\rm eff} &=& v^2 \left[  - \partial_\mu
 \theta \, \partial^\mu \theta + \frac{1}{4m^2} (\partial_\mu \theta \,
 \partial^\mu \theta)^2 - {1 \over 4 m^4} \; (\partial_\mu
 \theta \, \partial^\mu \theta)^3 \right.
 \\
 && \left. \qquad\qquad\qquad\qquad + {1 \over 4 m^4}
 \; (\partial_\mu \theta \, \partial^\mu \theta)
 \partial_\lambda \partial^\lambda
 (\partial_\nu \theta \, \partial^\nu \theta) + \cdots \right] \,,
 \nonumber
\end{eqnarray}
where the ellipses in ${\cal L}$ represent terms which are
suppressed by more than four inverse powers of $m$. The inverse
powers of $m$ which pre-multiply all of the interactions in this
Lagrangian are a consequence of the virtual $\hat\varphi_m$
exchanges which are required in order to produce them within the
full theory. The explicit numerical factors in each term are an
artifact of leading order perturbation theory, and receive
corrections order by order in $\lambda$. Computing 2-particle
scattering using this effective theory gives a result for which
the low-energy suppression by powers of $E/m$ are explicit due to
the derivative form of the interactions.

What is interesting about the lagrangian, eq.~(\ref{Leffform}),
for the present purposes is that the successive effective
couplings involve successively more powers of $1/m^2$. In
particular, this keeps them from having non-negative engineering
dimension and so makes the effective theory manifestly
non-renormalizable. If someone were to hand us this theory we
might therefore throw up our hands and conclude that we cannot
predictively compute quantum corrections. However in this case we
know this theory simply expresses the low-energy limit of a full
theory which {\it is} renormalizable, and so for which quantum
corrections can be explicitly computed. Why can't these
corrections also be expressed using the effective theory?

The answer is that they can, and this is by far the most efficient
way to compute these corrections to observables in the low-energy
limit where $E \ll m$. The key to computing these corrections is
to systematically exploit the low-energy expansion in powers of
$E/m$ which underlies using the action, eq.~(\ref{Leffform}) in
the first place.

\subsection{Computing loops}

To explore quantum effects consider evaluating loop graphs using
the toy-model effective lagrangian, which we may write in the
general form
\begin{equation} \label{genformLeff}
 {\cal L}_{\rm eff} = v^2 m^2 \sum_{id} {c_{id} \over m^d} \; {\cal
 O}_{id},
\end{equation}
where the sum is over interactions, ${\cal O}_{id}$, involving $i$
powers of the dimensionless field $\theta$ and $d$ derivatives.
The power of $m$ pre-multiplying each term is chosen to ensure
that the coefficient $c_{id}$ is dimensionless, and we have seen
that these coefficients are $O(1)$ at leading order in
$\lambda^2$. To be completely explicit, in the case of the
interaction ${\cal O} = (\partial_\mu \theta \, \partial^\mu
\theta)^2$ we have $i = d = 4$ and we found earlier that $c_{44} =
\frac14 + O(\lambda^2)$ for this term. Notice that Lorentz
invariance requires $d$ must be even, and the $U(1)$ symmetry
implies every factor of $\theta$ is differentiated at least once,
and so $d \ge i$. We may ignore all terms with $i = 1$ since these
are linear in $\partial_\mu\theta$ and so must be a total
derivative.\footnote{Terms like total derivatives, which do not
contribute to the observables of interest, are called {\it
redundant} and may be omitted when writing the effective
Lagrangian.} Furthermore, the only term with $i = 2$ is the
kinetic term, which we take as the unperturbed Lagrangian, and so
for the interactions we may restrict the sum to $i \ge 3$.

With these definitions it is straightforward to track the powers
of $v$ and $m$ that interactions of the form (\ref{genformLeff})
contribute to an $L$-loop contribution to a scattering amplitude,
${\cal A}(E)$, at centre-of-mass energy $E$. (The steps presented
here closely follow the discussion of refs.~\cite{CBLivRev,CBGB}.)
Imagine using this lagrangian to compute a contribution to the
scattering amplitude, ${\cal A}(E)$, coming from a Feynman graph
involving ${\cal E}$ external lines; $I$ internal lines and
$V_{ik}$ vertices. (The subscript $i$ here counts the number of
lines which converge at the vertex, while $k$ counts the power of
momentum which appears.) These constants are not all independent,
since they are related by the identity $2 I + E = \sum_{ik} i \,
V_{ik}$. It is also convenient to trade the number of internal
lines, $I$, for the number of loops, $L$, defined by $L = 1 + I -
\sum_{ik} V_{ik}$.

We now use dimensional analysis to estimate the result of
performing the integration over the internal momenta, using
dimensional regularization to regulate the ultraviolet
divergences. If all external momenta and energies are of order $E$
then the size of a dimensionally-regularized integral is given on
dimensional grounds by the appropriate power of $E$, we find
\begin{eqnarray} \label{PCresult1}
 {\cal A}(E) &\sim& v^2
 m^2 \left( {1 \over v} \right)^{\cal E}
 \left( { m \over 4 \pi v} \right)^{2L}
 \left( {E \over m} \right)^P \nonumber\\
 \label{PCresult2}
 &\sim& v^2
 E^2 \left( {1 \over v} \right)^{\cal E}
 \left( { E \over 4 \pi v} \right)^{2L} \prod_{i} \prod_{d>2}
 \left( {E \over m} \right)^{(d-2)V_{id}}
 \,.
\end{eqnarray}
where $P = 2 + 2L + \sum_{id} (d-2) V_{id}$. This is the main
result, since it shows which graphs contribute to any order in
$E/m$ using a nonrenormalizable theory.\footnote{It is here that
the convenience of dimensional regularization is clear, since it
avoids keeping track of powers of a cutoff like $\Lambda$, which
drops out of the final answer for an observable in any case.}

To see how eqs.~(\ref{PCresult1}) are used, consider the first few
powers of $E$ in the toy model. For any ${\cal E}$ the leading
contributions for small $E$ come from {\it tree} graphs, {\it
i.e.}~those having $L = 0$. The tree graphs that dominate are
those for which $\sum_{id}' (d-2)V_{id}$ takes the smallest
possible value. For example, for 2-particle scattering ${\cal E} =
4$ and so precisely one tree graph is possible for which
$\sum_{id}'(d-2)V_{id} = 2$, corresponding to $V_{44} = 1$ and all
other $V_{id} = 0$. This identifies the single graph which
dominates the 4-point function at low energies, and shows that the
resulting leading energy dependence in this case is ${\cal A}(E)
\sim E^4/(v^2 \, m^2)$, as was also found earlier in the full
theory. The numerical coefficient can be obtained in terms of the
effective couplings by more explicit evaluation of the appropriate
Feynman graph.

The next-to-leading behavior is also easily computed using the
same arguments. Order $E^6$ contributions are achieved if and only
if either: ({\it i}) $L = 1$ and $V_{i4} = 1$, with all others
zero; or ({\it ii}) $L = 0$ and $\sum_{i} \Bigl(4 V_{i6} + 2
V_{i4} \Bigr) = 4$. Since there are no $d=2$ interactions, no
one-loop graphs having 4 external lines can be built using
precisely one $d=4$ vertex and so only tree graphs can contribute.
Of these, the only two choices allowed by ${\cal E} = 4$ at order
$E^6$ are therefore the choices: $V_{46} = 1$, or $V_{34} = 2$.
Both of these contribute a result of order ${\cal A}(E) \sim
E^6/(v^2 \, m^4)$.

Besides showing how to use the effective theory to compute to any
order in $E/m$, eq.~(\ref{PCresult2}) also shows the domain of
approximation of the effective-theory calculation. The validity of
perturbation theory within the effective theory relies only on the
assumptions $E \ll 4 \pi v$ and $E \ll m$. In particular, it does
\emph{not} rely on the ratio $m/4\pi v = \lambda/4\pi$ being
small, even though there is a factor of this order appearing for
each loop. This factor does not count loops in the effective
theory because it is partially cancelled by another factor, $E/m$,
which also comes with every loop. $\lambda/4\pi$ {\it does} count
loops within the full theory, of course. This calculation simply
shows that the small-$\lambda$ approximation is only relevant for
predicting the values of the effective couplings, but are
irrelevant to the problem of computing the energetics of
scattering amplitudes given these couplings.

\subsection{The effective lagrangian logic}

These calculations show how to calculate predictively --- {\it
including loops} --- using a non-renormalizable effective theory.

\smallskip\noindent{\it Step I:} Choose the accuracy desired in
the answer ({\it e.g.} a 1\% accuracy might be desired.)

\smallskip\noindent{\it Step II:} Determine how many powers of
$E/m$ are required in order to achieve the desired accuracy.

\smallskip\noindent{\it Step III:} Use a calculation like the
one above to identify which effective couplings in ${\cal L}_{\rm
eff}$ can contribute to the observable of interest to the desired
order in $E/m$. This always requires only a finite number (say:
$N$) of terms in ${\cal L}_{\rm eff}$ to any finite accuracy.

There are two alternative versions of the fourth and final step,
depending on whether or not the underlying microscopic theory ---
like the $\phi$ theory in the toy model --- is known.

\smallskip\noindent{\it Step IV-A:} If the underlying theory is
known and calculable, then compute the required coefficients of
the $N$ required effective interactions to the accuracy required.
Alternatively,

\smallskip\noindent{\it Step IV-B:} If the underlying theory is
unknown, or is too complicated to permit the calculation of ${\cal
L}_{\rm eff}$, then leave the $N$ required coefficients as free
parameters. The procedure is nevertheless predictive if more than
$N$ observables can be identified whose predictions depend only on
these parameters.

The effective lagrangian is in this way seen to be predictive even
though it is not renormalizable in the usual sense. Renormalizable
theories are simply the special case of Step {\it IV-B} where one
stops at zeroeth order in $E/m$, and so are the ones which
dominate in the limit that the light and heavy scales are very
widely separated. In fact, this is {\it why} renormalizable
interactions are so important when describing Nature.

The success of the above approach is well-established in many
areas outside of gravitational physics, with non-renormalizability
being the signal that one is seeing the virtual effects due to
some sort of heavier physics. Historically, one of earliest
examples known was the non-renormalizable interactions of chiral
perturbation theory which describe well the low-energy scattering
of pions, kaons and nucleons. It is noteworthy that this success
requires the inclusion of the loop corrections within this
effective theory. The heavier physics in this case is the
confining physics of the quarks and gluons from which these
particles are built, and whose complicated dynamics has so far
precluded calculating the effective couplings from first
principles. The effective theory works so long as one restricts to
center-of-mass energies smaller than roughly 1 GeV.

The Fermi (or $V-A$) theory of the weak interactions is a similar
example, which describes the effects of virtual $W$-boson exchange
at energies well below the $W$-boson mass, $M_{\scriptscriptstyle
W} = 80$ GeV. This theory provides an efficient description of the
low-energy experiments, with an effective coupling,
$G_{\scriptstyle F}/\sqrt2 = g^2/8M_{\scriptscriptstyle W}^2$
which in this case is calculable in terms of the mass and
coupling, $g$, of the $W$ boson. In this case agreement with the
precision of the measurements again requires the inclusion of
loops within the effective theory.

\section{Gravity as an effective theory}

Given the previous discussion of of the toy model, it is time to
return to the real application of interest for this chapter:
General Relativity. The goal is to be able to describe
quantitatively quantum processes in GR, and to be able to compute
the size of quantum corrections to the classical processes on
which the tests of GR are founded.

Historically, the main obstacle to this program has been that GR
is not renormalizable, as might be expected given that its
coupling (Newton's constant), $G = (8 \pi M_p^2)^{-1}$, has
engineering dimension (mass)${}^{-2}$ in units where $\hbar = c=
1$. But we have seen that non-renormalizable theories can be
predictive in much the same way as are renormalizable ones,
provided that they are interpreted as being the low-energy limit
of some more fundamental microscopic theory. For gravity, this
more microscopic theory is as yet unknown, although these pages
contain several proposals for what it might be. Happily, as we
have seen for the toy model, their effective use at low energies
does not require knowledge of whatever this microscopic theory
might be. In this section the goal is to identify more thoroughly
what the precise form of the low-energy theory really is for
gravity, as well as to identify what the scales are above which
the effective theory should not be applied.

\subsection{The effective action}

For GR the low-energy fields consist of the metric itself,
$g_{\mu\nu}$. Furthermore, since we do not know what the
underlying, more microscopic theory is, we cannot hope to compute
the effective theory from first principles. Experience with the
toy model of the previous section instead suggests we should
construct the most general effective Lagrangian which is built
from the metric and organize it into a derivative expansion, with
the terms with the fewest derivatives being expected to dominate
at low energies. Furthermore we must keep only those effective
interactions which are consistent with the symmetries of the
problem, which for gravity we can take to be general covariance.

These considerations lead us to expect that the Einstein-Hilbert
action of GR should be considered to be just one term in an
expansion of the action in terms of derivatives of the metric
tensor. General covariance requires this to be written in terms of
powers of the curvature tensor and its covariant derivatives,
\begin{eqnarray}
\label{gravaction}
 - \, {{\cal L}_{\rm eff} \over \sqrt{- g}} &=& \lambda
+ \frac{M_p^2}{2}  \, R + a_1 \, R_{\mu\nu} \, R^{\mu\nu} + a_2 \,
R^2 +  a_3 \, R_{\mu\nu\lambda\rho} R^{\mu\nu\lambda\rho} + a_4 \,
\Box R \nonumber\\
&&\quad + {b_1 \over m^2}\; R^3 + {b_2 \over m^2} \; R R_{\mu\nu}
R^{\mu\nu} + {b_3\over m^2} \; R_{\mu\nu} R^{\nu\lambda}
{R_\lambda}^\mu + \cdots \,,
\end{eqnarray}
where ${R^\mu}_{\nu\lambda\rho}$ is the metric's Riemann tensor,
$R_{\mu\nu} = {R^\lambda}_{\mu\lambda\nu}$ is its Ricci tensor,
and $R = g^{\mu\nu}R_{\mu\nu}$ is the Ricci scalar, each of which
involves precisely two derivatives of the metric.

The first term in eq.~(\ref{gravaction}) is the cosmological
constant, which is dropped in what follows since observations
imply $\lambda$ is (for some reason) extremely small. Once this is
done the leading term in the derivative expansion is the
Einstein-Hilbert action whose coefficient, $M_p \sim 10^{18}$ GeV,
has dimensions of squared mass, whose value defines Newton's
constant. This is followed by curvature-squared terms having
dimensionless effective couplings, $a_i$, and curvature-cubed
terms with couplings inversely proportional to a mass, $b_i/m^2$,
(not all of which are written in eq.~(\ref{gravaction})). Although
the numerical value of $M_p$ is known, the mass scale $m$
appearing in the curvature-cubed (and higher) terms is not. But
since it appears in the denominator it is the lowest mass scale
which has been integrated out which should be expected to
dominate. For this reason $m$ is unlikely to be $M_p$, and one
might reasonably use the electron mass, $m_e = 5\times 10^{-4}$
GeV, or neutrino masses, $m_\nu \gsim 10^{-11}$ GeV, when
considering applications over the distances relevant in
astrophysics.

Experience with the toy model shows that not all of the
interactions in the lagrangian (\ref{gravaction}) need contribute
independently (or at all) to physical observables. For instance,
for most applications we may drop total derivatives (like $\Box
R$), as well as those terms which can be eliminated by performing
judicious field redefinitions \cite{CBLivRev}. Since the existence
of these terms does not affect the arguments about to be made, we
do not bother to identify and drop these terms explicitly here.

\subsection{Power counting}

Of all of the terms in the effective action, only the
Einstein-Hilbert term is familiar from applications of classical
GR. Although we expect naively that this should dominate at low
energies (since it involves the fewest derivatives), we now make
this more precise by identifying which interactions contribute to
which order in a low-energy expansion. We do so by considering the
low-energy scattering of weak gravitational waves about flat
space, and by repeating the power-counting exercise performed
above for the toy model to keep track of how different effective
couplings contribute. In this way we can see how the scales $M_p$
and $m$ enter into observables.

In order to perform this power counting we expand the above action
flat space, trading the full metric for a canonically normalized
fluctuation: $g_{\mu\nu} = \eta_{\mu\nu} + 2h_{\mu\nu}/M_p$. For
the present purposes what is important is that the expansion of
the curvature tensor (and its Ricci contractions) produces terms
involving all possible powers of $h_{\mu\nu}$, with each term
involving precisely two derivatives. Proceeding as before gives an
estimate for the leading energy-dependence of an $L$-loop
contribution to the a scattering amplitude, ${\cal A}$, which
involves ${\cal E}$ external lines and $V_{id}$ vertices involving
$d$ derivatives and $i$ attached graviton lines. (The main
difference from the previous section's analysis is the appearance
here of interactions involving two derivatives, coming from the
Einstein-Hilbert term.)

This leads to the estimate:
\begin{equation}
\label{GRcount1}
 {\cal A}(E) \sim m^2 M_p^2 \left( {1
 \over M_p} \right)^{\cal E} \left( {m \over 4 \pi M_p} \right)^{2
 L} \left( {m^2 \over M_p^2} \right)^{Z} \left( {E \over m}
 \right)^P
\end{equation}
where $Z = \sum_{id}' V_{id}$ and $P = 2 + 2L + \sum_{id}' (d-2)
V_{id}$. The prime on both of these sums indicates the omission of
the case $d=2$ from the sum over $d$. Grouping instead the terms
involving powers of $L$ and $V_{ik}$, eq.~(\ref{GRcount1}) becomes
\begin{equation}
\label{GRcount1a}
 {\cal A}_{\cal E}(E) \sim E^2 M_p^2 \left( {1
 \over M_p} \right)^{\cal E}
 \left( {E \over 4 \pi M_p} \right)^{2
 L} {\prod_{i} \prod_{d>2}} \left[{E^2 \over M_p^2}
 \left( {E \over m} \right)^{(d-4)}  \right]^{V_{id}} \,.
\end{equation}
Notice that no negative powers of $E$ appear here because $d$ is
even and because of the condition $d > 2$ in the product.

This last expression is the result we seek because it is what
shows how to make systematic quantum predictions for graviton
scattering. It does so by showing that the predictions of the full
gravitational effective lagrangian (involving all powers of
curvatures) can be organized into powers of $E/M_p$ and $E/m$, and
so we can hope to make sensible predictions provided that both of
these two quantities are small. Furthermore, all of the
corrections involve powers of $(E/M_p)^2$ and/or $(E/m)^2$,
implying that they may be expected to be {\it extremely} small for
any applications for which $E \ll m$. For instance, notice that
even if $E/m \sim 1$ then $(E/M_p)^2 \sim 10^{-42}$ if $m$ is
taken to be the electron mass. (Notice that factors of the larger
parameter $E/m$ do not arise until curvature-cubed interactions
are important, and this first occurs at subleading order in
$E/M_p$.)

Furthermore, it shows in detail what we were in any case inclined
to believe: that classical General Relativity governs the dominant
low-energy dynamics of gravitational waves. This can be seen by
asking which graphs are least suppressed by these small energy
ratios, which turns out to be those for which $L = 0$ and $P = 2$.
That is, arbitrary tree graphs constructed purely from the
Einstein-Hilbert action --- precisely the predictions of classical
General Relativity. For instance, for 2-graviton scattering we
have ${\cal E} = 4$, and so the above arguments predict the
dominant energy-dependence to be ${\cal A}(E) \propto (E/M_p)^2 +
\cdots$. This is borne out by explicit tree-level calculations
\cite{DeWitt} for graviton scattering, which give:
\begin{equation}
 {\cal A}_{\rm tree} = 8 \pi i G \,\left(
 \frac{s^3}{tu} \right)\,,
\end{equation}
for an appropriate choice of graviton polarizations. Here $s = -
(p_1 + p_2)^2$, $t = (p_1 - p_1')^2$ and $u = (p_1 - p_2')^2$ are
the usual Lorentz-invariant Mandelstam variables built from the
initial and final particle four momenta, all of which are
proportional to $E^2$. This shows both that ${\cal A} \sim
(E/M_p)^2$ to leading order, and that it is the physical,
invariant centre-of-mass energy, $E$, which is the relevant energy
for the power-counting analysis.

But the real beauty of a result like eq.~(\ref{GRcount1a}) is that
it also identifies which graphs give the subdominant corrections
to classical GR. The leading such a correction arises in one of
two ways: either ({\it i}) $L = 1$ and $V_{id} = 0$ for any $d\ne
2$; or ({\it ii}) $L = 0$, $\sum_i V_{i4} = 1$, $V_{i2}$ is
arbitrary, and all other $V_{id}$ vanish. That is, compute the
one-loop corrections using only Einstein gravity; or instead work
to tree level and include precisely one vertex from one of the
curvature-squared interactions in addition to any number of
interactions from the Einstein-Hilbert term. Both are suppressed
compared to the leading term by a factor of $(E/M_p)^2$, and the
one-loop contribution carries an additional factor of
$(1/4\pi)^2$. This (plus logarithmic complications due to infrared
divergences) are also borne out by explicit one-loop calculations
\cite{loopgraviton,IRcancel,DonoghueIR}. Although the use of
curvature-squared terms potentially introduces additional
effective couplings into the results,\footnote{For graviton
scattering in 4D with no matter no new couplings enter in this way
because all of the curvature-squared interactions turn out to be
redundant. By contrast, one new coupling turns out to arise
describing a contact interaction when computing the sub-leading
corrections to fields sourced by point masses.} useful predictions
can nonetheless be made provided more observables are examined
than there are free parameters.

Although conceptually instructive, calculating graviton scattering
is at this point a purely academic exercise, and is likely to
remain so until gravitational waves are eventually detected and
their properties are measured in detail. In practice it is of more
pressing interest to obtain these power-counting estimates for
observables which are of more direct interest for precision
measurements of GR, such as within the solar system. It happens
that the extension to these kinds of observables is often not
straightforward (and in some cases has not yet been done in a
completely systematic way), because they involve non-relativistic
sources (like planets and stars). Non-relativistic sources
considerably complicate the above power-counting arguments because
they introduce a new dimensionless small quantity, $v^2/c^2$,
whose dependence is not properly captured by the simple
dimensional arguments given above \cite{DonoghuePC}.

Nevertheless the leading corrections have been computed for some
kinds of non-relativistic sources in asymptotically-flat
spacetimes \cite{DonoghueCalcs,GoldbergerRothstein}. These show
that while relativistic corrections to the observables situated a
distance $r$ away from a gravitating mass $M$ are of order
$GM/rc^2$, the leading quantum corrections are suppressed by
powers of the much smaller quantity $G\hbar/r^2c^3$. For instance,
while on the surface of the Sun relativistic corrections are of
order $GM_\odot/R_\odot c^2 \sim 10^{-6}$, quantum corrections are
completely negligible, being of order $G \hbar/R_\odot^2 c^3 \sim
10^{-88}$. Clearly the classical approximation to GR is {\it
extremely} good for solar-system applications.

Another important limitation to the discussion as given above is
its restriction to perturbations about flat space. After all,
quantum effects are also of interest for small fluctuations about
other spacetimes. In particular, quantum fluctuations generated
during a past epoch of cosmological inflationary expansion appear
to provide a good description of the observed properties of the
cosmic microwave background radiation. Similarly, phenomena like
Hawking radiation rely on quantum effects near black holes, and
the many foundational questions these raise have stimulated their
extensive theoretical study, even though these studies may not
lead in the near term to observational consequences. Both black
holes and cosmology provide regimes for which detailed quantum
gravitational predictions are of interest, but for which
perturbations about flat space need not directly apply.

A proper power-counting of the size of quantum corrections is also
possible for these kinds of spacetimes by examining perturbations
about the relevant cosmological or black-hole geometry, although
in these situations momentum-space techniques are often less
useful. Position-space methods, like operator-product expansions,
can then provide useful alternatives, although as of this writing
comparatively few explicit power-counting calculations have been
done using these. The interested reader is referred to the longer
review, \cite{CBLivRev}, for more discussion of this, as well as
of related questions which arise concerning the use of effective
field theories within time-dependent backgrounds and in the
presence of event horizons.

\section{Summary}

General Relativity provides a detailed quantitative description of
gravitational experiments in terms of a field theory which is not
renormalizable. It is the purpose of the present article to
underline the observation that gravity is not the only area of
physics for which a non-renormalizable theory is found to provide
a good description of experimental observations, and we should use
this information to guide our understanding of what the limits to
validity might be to its use.

The lesson from other areas of physics is clear: the success of a
non-renormalizable theory points to the existence of a new
short-distance scale whose physics is partially relevant to the
observations of interest. What makes this problematic for
understanding the theory's quantum predictions is that it is often
the case that we do often not understand what the relevant new
physics is, and so its effects must be parameterized in terms of
numerous unknown effective couplings. How can predictions be made
in such a situation?

What makes predictions possible is the observation that only
comparatively few of these unknown couplings are important at low
energies (or long distances), and so only a finite number of them
enter into predictions at any fixed level of accuracy. Predictions
remain possible so long as more observables are computed than
there are parameters, but explicit progress relies on being able
to identify which of the parameters enter into predictions to any
given degree of precision.

In the previous pages it is shown how this identification can be
made for the comparatively simple case of graviton scattering in
flat space, for which case the size of the contribution from any
given effective coupling can be explicitly estimated. The central
tool is a power-counting estimate which tracks the power of energy
which enters into any given Feynman graph, and which duplicates
for GR the similar estimates which are made in other areas of
physics. The result shows how General Relativity emerges is the
leading contribution to an effective theory of some more
fundamental picture, with its classical contributions being shown
to be the dominant ones, but with computable corrections which can
be explicitly evaluated in a systematic expansion to any given
order in a low-energy expansion. This shows how a theory's
non-renormalizability need not preclude its use for making
sensible quantum predictions, provided these are made only for low
energies and long distances.

This kind of picture is satisfying because it emphasizes the
similarity between many of the problems which are encountered in
GR and in other areas of physics. It is also conceptually
important because it provides control over the size of the
theoretical errors which quantum effects would introduce into the
classical calculations against which precision measurements of
General Relativity are compared. These estimates show that the
errors associated with ignoring quantum effects is negligible for
the systems of practical interest.

There is a sense for which this success is mundane, in that it
largely confirms our prejudices as to the expected size of quantum
effects for macroscopic systems based purely on dimensional
analysis performed by building dimensionless quantities out of the
relevant parameters like $G$, $\hbar$, $c$, $M$ and $R$. However
the power-counting result is much more powerful: it identifies
which Feynman graphs contribute at any given power of energy, and
so permits the detailed calculation of observables as part of a
systematic low-energy expansion.

It is certainly true the small size of quantum contributions in
the solar system in no way reduces the fundamental mysteries
described elsewhere in these pages that must be resolved in order
to properly understand quantum gravity at fundamentally small
distances. However it is important to understand that these
problems are associated with small distance scales and not with
large ones, since this focusses the discussion as to what is
possible and what is not when entertaining modifications to GR. In
particular, although it shows that we are comparatively free to
modify gravity at short distances without ruining our
understanding of gravitational physics within the solar system, it
also shows that we are not similarly protected from {\it
long}-distance modifications to GR.

This observation is consistent with long experience, which shows
that it is notoriously difficult to modify GR at long distances in
a way which does not introduce unacceptable problems such as
various sorts of instabilities to the vacuum. Such
vacuum-stability problems are often simply ignored in some circles
on the grounds that `quantum gravity' is not yet understood, in
the hope that once it is it will somehow also fix the stability
issues. However our ability to quantify the size of low-energy
quantum effects in gravity shows that we need not await a more
complete understanding of gravity at high energies in order to
make accurate predictions at low energies. And since the vacuum is
the lowest-energy state there is, we cannot expect unknown
short-distance physics to be able to save us from long-distance
sicknesses.

Calculability at low energies is a double-edged sword. It allows
us to understand why precision comparison between GR and
experiment is possible in the solar system, but it equally forces
us to reject alternative theories which have low-energy problems
(like instabilities) as being inadequate.

\section*{Acknowledgements}

I thank Daniel Oriti for the kind invitation to contribute to this
volume, and for his subsequent patience. My understanding of this
topic was learned from Steven Weinberg, who had been making the
points made here for decades before my arrival on the scene. My
research on these and related topics is funded by the Natural
Sciences and Engineering Research Council of Canada, as well as by
funds from McMaster University and the Killam Foundation.





\begin{thereferences}{widest citation in source list}

\bibitem{CWLivRev}
%
C.M. Will, {\it Living Rev. Rel.} {\bf 4} (2001) 4
(gr-qc/0103036).

\bibitem{CBLivRev}
%
C.P. Burgess, {\it Living Rev. Rel.} {\bf 7} (2004) 5
(gr-qc/0311082).

\bibitem{ChiPT}
S. Weinberg, {\it Physical Review Letters} {\bf 18} (1967) 188;
{\it Physical Review} {\bf 166} (1968) 1568;
%
C.G. Callan, S. Coleman, J. Wess and B. Zumino, {\it Physical
Review} {\bf 177} (1969) 2247;
%
S. Weinberg, {\it Physica} {\bf 96A} (1979) 327--340;
%
J. Gasser and H. Leutwyler, {\it Annals of Physics} (NY) {\bf 158}
(1984) 142.

\bibitem{CBGB}
C.P. Burgess, {\sl Goldstone and Pseudo-Goldstone Bosons in
Nuclear, Particle and Condensed-Matter Physics}, {\it Physics
Reports} {\bf 330} (2000) 193--261, ({hep-ph/9808176}).

\bibitem{DeWitt}
B.S.~DeWitt, {\it Phys. Rev.} {\bf 162} (1967) 1239.

\bibitem{loopgraviton}
D.C.~Dunbar and P.S.~Norridge, {\it Nucl. Phys.} {\bf B433} (1995)
181.

\bibitem{IRcancel}
S.~Weinberg, {\it Phys. Rev.} {\bf 140} (1965) 516.

\bibitem{DonoghueIR}
J.F.~Donoghue and T.~Torma, {\it Phys. Rev.} {\bf D60} (1999)
024003 (hep-th/9901156).

\bibitem{DonoghuePC}
J.F.~Donoghue and T.~Torma, {\it Phys. Rev.} {\bf D54} (1996)
4963-4972 ({\tt hep-th/9602121}).

\bibitem{DonoghueCalcs}
%
J.F.~Donoghue, {\it Phys. Rev. Lett.} {\bf 72} (1994) 2996 -- 2999
(gr-qc/9310024);
%
J.F.~Donoghue, {\it Phys. Rev.} {\bf D50} (1994) 3874--3888 ({\tt
gr-qc/9405057});
%
N.E.J.~Bjerrum-Borh, J.F.~Donoghue and B.R.~Holstein, {\it Phys.
Rev.} {\bf D68} (2003) 084005, Erratum-ibid. {\bf D71} (2005)
069904 (hep-th/0211071);
%
J.F.~Donoghue, B.R.~Holstein, B.~Garbrecht and T.~Konstandin, {\it
Phys. Lett.} {\bf B529} (2002) 132--142 (hep-th/0112237).


\bibitem{GoldbergerRothstein}
W.D. Goldberger and I.Z. Rothstein (hep-th/0409156).
\end{thereferences}
\end{document}